\PassOptionsToPackage{table,xcdraw}{xcolor}

\documentclass[sigconf, natbib=true]{acmart}

\AtBeginDocument{%
  \providecommand\BibTeX{{%
    \normalfont B\kern-0.5em{\scshape i\kern-0.25em b}\kern-0.8em\TeX}}}

\usepackage{caption}
\usepackage{subcaption}
\usepackage{lscape}
\usepackage{soul}

\usepackage{booktabs}
\usepackage[table,xcdraw]{xcolor}

\copyrightyear{2023}
\acmYear{2023}
\setcopyright{rightsretained}
\acmConference[CHIIR'23]{ACM SIGIR Conference on Human Information Interaction and Retrieval}{March 19--23, 2023}{Austin, TX, USA}
\acmBooktitle{ACM SIGIR Conference on Human Information Interaction and Retrieval (CHIIR'23), March 19--23, 2023, Austin, TX, USA}
\acmDOI{10.1145/3576840.3578307}
\acmISBN{979-8-4007-0035-4/23/03}

\begin{document}

\title[From 10 Blue Links Pages to Feature-Full Search Engine Results Pages]{From 10 Blue Links Pages to Feature-Full Search Engine Results Pages - Analysis of the Temporal Evolution of SERP Features}



\author{Bruno Oliveira}
\affiliation{%
  \institution{Faculty of Engineering of the University of Porto}
  \country{Porto, Portugal}}
\email{up201605516@edu.fe.up.pt}

\author{Carla Teixeira Lopes}
\affiliation{%
  \institution{Faculty of Engineering of the University of Porto and INESC-TEC}
  \country{Porto, Portugal}}
\email{ctl@fe.up.pt}

\begin{abstract}
Web Search Engine Results Pages (SERP) are one of the most well-known and used web pages. These pages have started as simple ``10 blue links'' pages, but the information in SERP currently goes way beyond these links. Several features have been included in these pages to complement organic and sponsored results and attempt to provide answers to the query instead of just pointing to websites that might deliver that information. In this work, we analyze the appearance and evolution of SERP features in the two leading web search engines, Google Search and Microsoft Bing. Using a sample of SERP from the Internet Archive, we analyzed the appearance and evolution of these features. We found that SERP are becoming more diverse in terms of elements, aggregating content from different verticals and including more features that provide direct answers.
\end{abstract}



\maketitle

\section{Introduction}
Who hasn't used a web search engine to look online for health information? Nowadays, 5,900,000 Google searches are conducted per minute, the 4th most popular activity on the Internet~\cite{MediaUsage}. These statistics imply that web search user interfaces are also at the top of the most used ones. Moreover, as user experience influences the perception of relevance, these interfaces are gaining importance ~\cite{Baeza-yates2011}. 

Search Engine Results Pages (SERP) are an essential component of web search user interfaces. These pages display retrieval results as vertical lists and have started as simple ``10 blue links'' pages. Although search engines always have presented results in a relatively consistent format, SERP have evolved and started to include additional information in the form of SERP features. 

Although some works provide an in-depth analysis of search user interfaces~\cite{10.5555/1631268}, the temporal evolution of SERP is understudied. Besides contributing to preserving the history of web search user interfaces, an evolutionary analysis provides insights into how content, layout, and navigation evolved, which can be used for further and deeper analysis. In another work, we analyzed the overall development of Google SERP with a greater focus on visual identity, navigation, user inputs, and organic and sponsored results~\cite{serpevolutionFull}. Here, we focus solely on SERP features, doing an evolutionary analysis of the two most popular web search engines, Google and Microsoft Bing. For this analysis, we have used 5,653 Google SERP and 2,267 Bing SERP captured from the Internet Archive.

This work has two significant contributions. First, we systematize the features that appear or have appeared in a SERP, defining each and providing visual examples. This systematization can be helpful in future studies with the SERP as their focus and contributes to establishing a common terminology. Second, we analyze the evolution of each SERP feature.

\section{Related work} \label{sec:relwork}
There are whole books and monographs dedicated to this subject~\cite{10.5555/1631268, 10.5555/2502704, 10.5555/1824082, 6812556, Wilson2011, Bierig2021}. Despite such research, given our focus, we only describe works that analyze SERP features. 

Moran and Goray~\cite{Moran2019} studied the anatomy of SERP, defining the terminology for SERP elements. Nielsen Norman Group uses this terminology in several articles~\cite{ThreeKeySERP,Moran2019,FeiFei20,MoranAbandon}. In their `Search Patterns' book, Morville and Callender~\cite{10.5555/1824082}, apart from addressing the anatomy of the search process and related behavior, also list elements and principles of interaction design, illustrating many user interface design patterns around search websites.
To the best of our knowledge, no works systematically analyze SERP features over time.

\section{Methodology} \label{sec:method}

Google Search currently has 91.4\% of the market share~\cite{Chris2020}, a leadership that goes back to 2002~\cite{MktShare20102022,MktShare20002013}, while Microsoft Bing is second-ranked with 3.1\%. In this context, we decided to focus our analysis on these two search engines, responsible for 95\% of the worldwide market share. For this analysis, we built a sample of Google and Bing desktop SERP interfaces over time. The Internet Archive keeps snapshots and the respective HTML version of web pages over time. Its collection contains 588 billion web pages~\cite{InternetArchive1996}. Using Internet Archive's \emph{Wayback CDX Server API} and the Google Trends most searched queries in each year of analysis\footnote{Available at \url{https://bedgarone.github.io/serpevolution/mostsearchedqueries}}, we retrieved 5.653 captures from Google and 2.267 captures from Bing\footnote{Available at \url{https://doi.org/10.25747/991g-f765}}. We chose head queries for their greater capacity to trigger features in SERP, the focus of our analysis.

\begin{figure}[h]
    \centering
    \includegraphics[width=0.7\columnwidth]{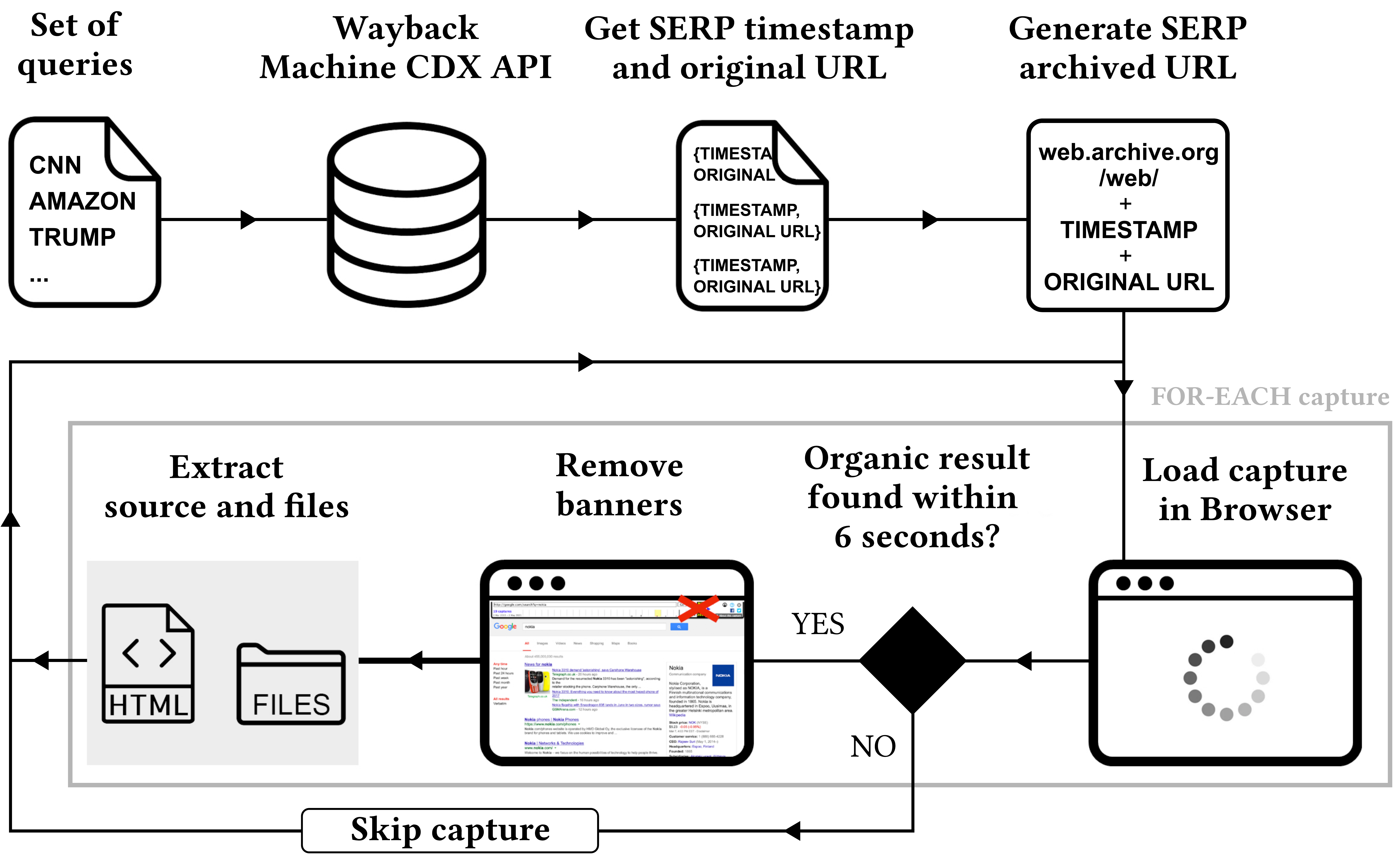}
    \caption{Extracting captures procedure}
    \label{fig:diagram}
\end{figure}

We used Python and Selenium Webdriver to visit each capture online, check if the capture was valid, save the HTML version, and generate a screenshot. The capture process is shown in Figure~\ref{fig:diagram}. The \emph{original URL} is the URL of the original SERP (e.g., \url{google.com/search?q=photography}), while the \emph{archived URL} is the URL of its archived version (e.g., \url{web.archive.org/web/20160125203434/www.google.com/search?q=photography}). The process concludes with generating full-height screenshots of every HTML version opened in another browser instance in headless mode. We produced screenshots considering the most popular screen size at the time of the capture, as stated by the statistics~\cite{TeoalidaScreen}.

\begin{figure}[h]
    \centering
    \includegraphics[width=0.8\columnwidth]{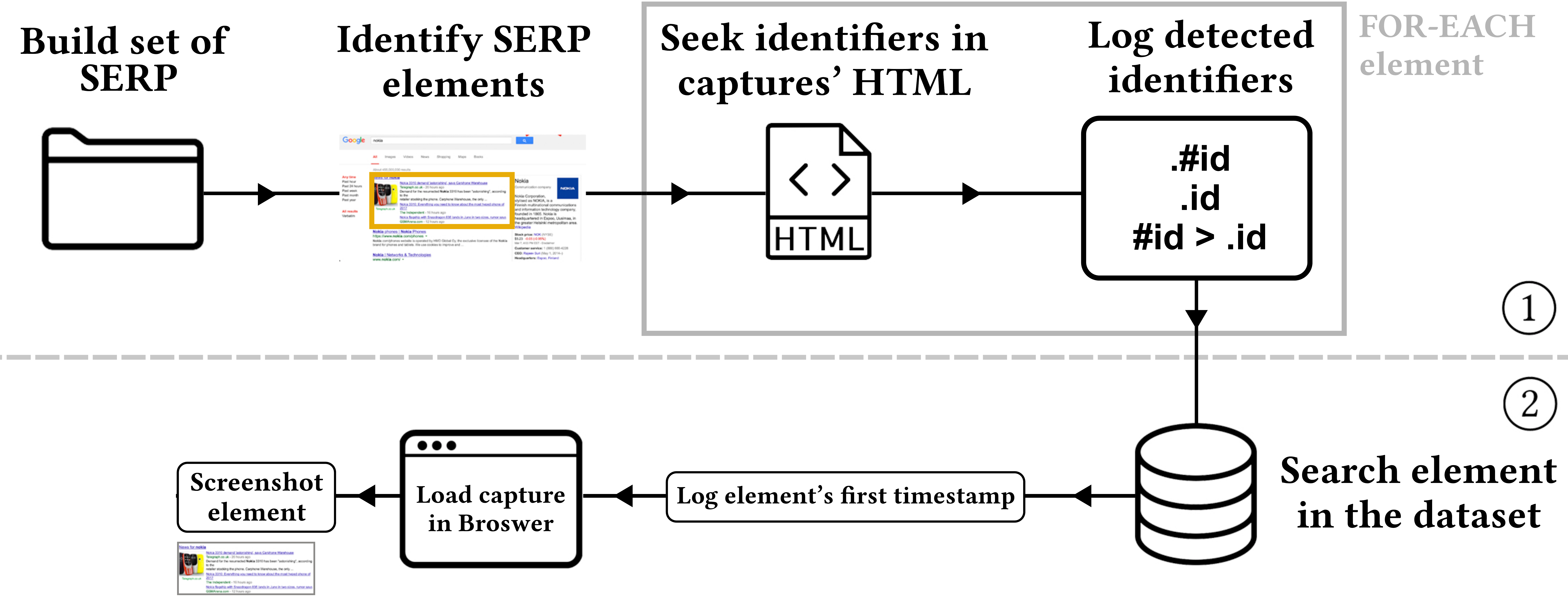}
    \caption{Detection and analysis of elements procedure}
    \label{fig:diagramanaly}
\end{figure}

The analysis process included two main stages, as shown in Figure~\ref{fig:diagramanaly}. First, we have extracted a sample of captures from the primary dataset to identify SERP elements. For each month with captures in the primary dataset, we manually looked at the screenshots of that month's captures and selected the capture with the most features. In the end, this set included 117 captures, with which we visually identified SERP elements. We analyzed each element's source code, looking for identifiers to locate the element in a later automated process. Element identifiers consist of HTML classes, ids, tags, or a combination of these using CSS selectors (e.g., \emph{`.knowledge-panel'}, \emph{`\#tads'} or \emph{`\#newsbox'}). All the encountered identifiers were logged and listed on the website\footnote{Available at \url{https://bedgarone.github.io/serpevolution/elements}}.

In the second stage, we automated the detection of these features over time, allowing the exploration of a more significant number of cases. Finding a feature with these identifiers triggers a function that stores the date of the element's appearance in a log file. The function also receives the element's upper-left corner coordinates, width, and height, generating and saving its image in the element's folder.

A more detailed description of the methodology can be found in another paper~\cite{serpevolutionFull} related to this work.

\section{Features Analysis} \label{sec:componentsevol}
\begin{figure*}[h]
    \centering
    \includegraphics[width=1.0\textwidth]{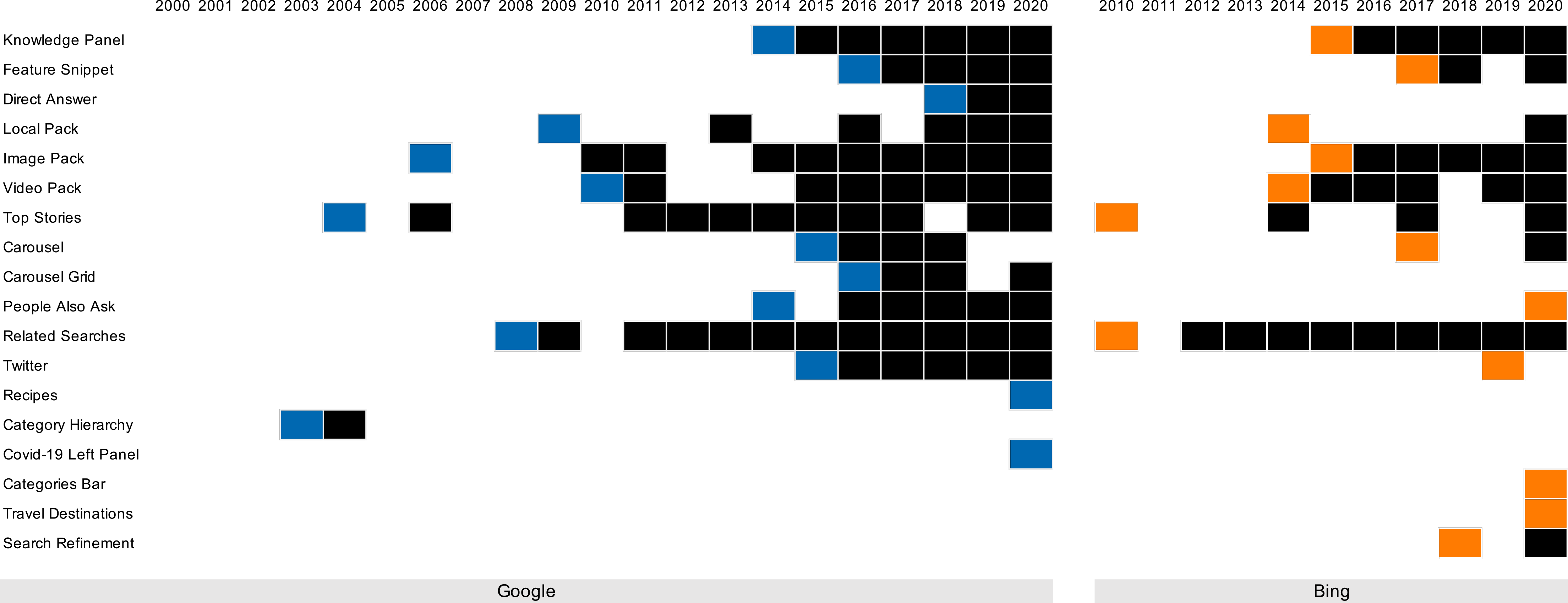}
    \caption{Detected presence of Google (left) and Bing (right) SERP features over time. Colored cells mark feature release years.}
    \label{fig:presence}
\end{figure*}

SERP features complement organic and sponsored results, attempting to provide answers to the query without just pointing to websites that might deliver that information. In this section, we describe the features that we identified in the collected SERP. Most elements appear in both search engines, but some are exclusive, appearing in only one. Figure~\ref{fig:presence} shows the features and their presence period in the search engines, while release years are marked with colored cells.

\subsection{Features existing in Google and Bing}
Given the similarity of the \emph{Image Pack}, already described in another paper \cite{serpevolutionFull}, and the \emph{Video Pack}, we will only consider the latter. 

\begin{figure}[H]
    \centering
    \includegraphics[width=1.0\columnwidth]{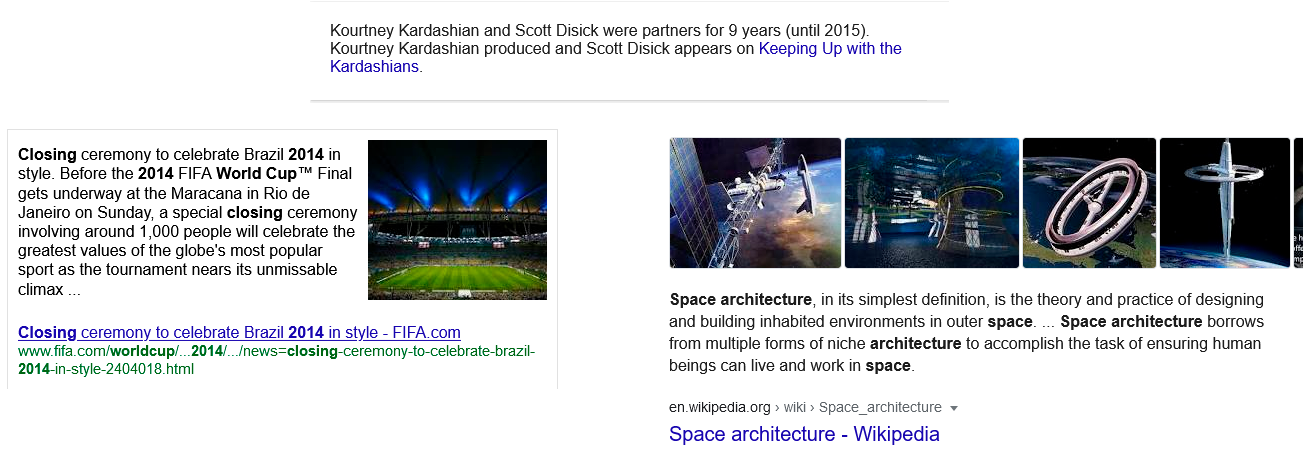}
    \caption{Google's Featured Snippets from 2016 (top), 2017 (left) and 2020 (right)}
    \label{fig:fsnippets}
\end{figure}

\begin{figure}[h]
    \centering
    \includegraphics[width=1.0\columnwidth]{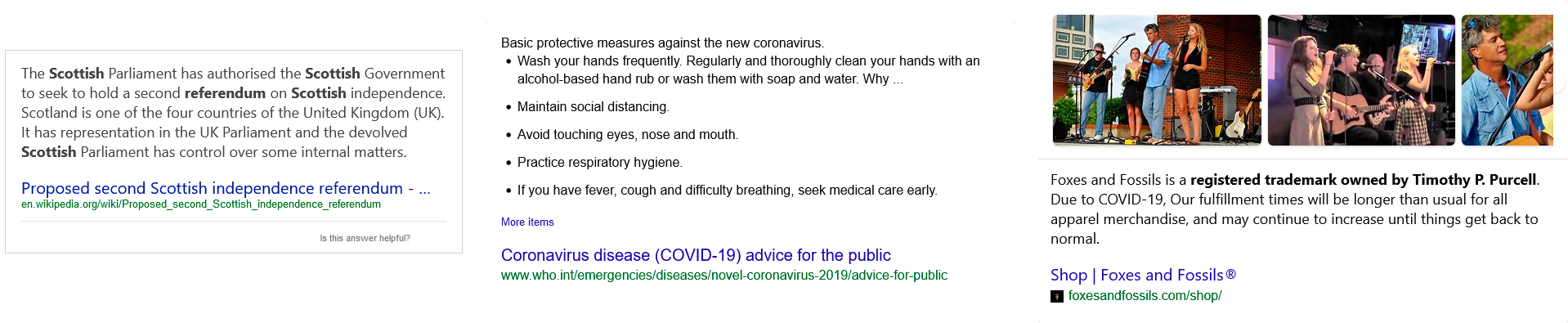}
    \caption{Bing's Featured Snippets from 2017 (left) and 2020 (center and right)}
    \label{fig:bingfsnipt}
\end{figure}

The \textbf{Featured Snippets}, seen in Figures \ref{fig:fsnippets} and \ref{fig:bingfsnipt}, are answer boxes in which Google responds to a question-related query based on information taken from a page \cite{Vaughn2019}. This element appeared for the first time in 2016, lasting until now. Its positioning is consistently at the top of the results container. Initially, a featured snippet contained a short paragraph with answering information. It evolved to a general layout, used until now, consisting of a larger paragraph, a thumbnail at the upper-right corner, and the title and link to the information's source, to where it is possible to navigate. The answer also started to be returned as an ordered list or table. Bing launched a simpler version of this layout with paragraph and source only. In Google, instead of an image, a video or a carousel of images may also accompany it. In 2018, when available, Google introduced the date of the source's publication after the information paragraph. In Bing, from 2020, the paragraph is just one of the options, as the content can now be displayed in a bullet list or accompanied by a carousel of images. Studies have found that these snippets help make more accurate decisions when containing the correct information~\cite{Bink2022}.

\begin{figure}[h]
  \centering
    \includegraphics[width=\columnwidth]{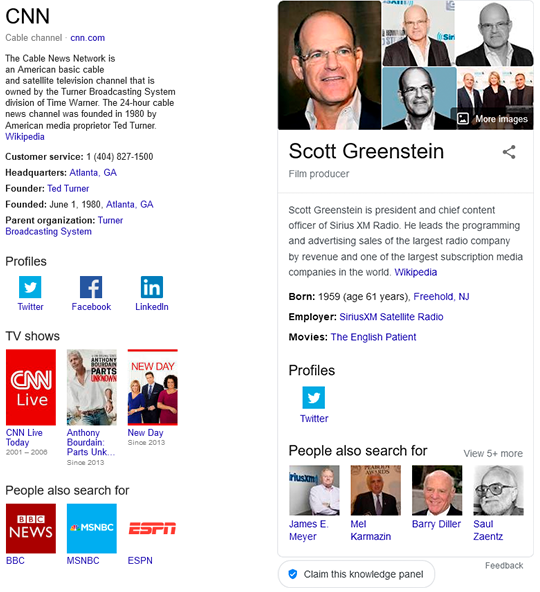}
    \caption{Google's Knowledge Panel from 2016 (left) and 2018 (right)}
    \label{fig:kpgoogle}
 \end{figure}

\begin{figure}[h]
    \includegraphics[width=\columnwidth]{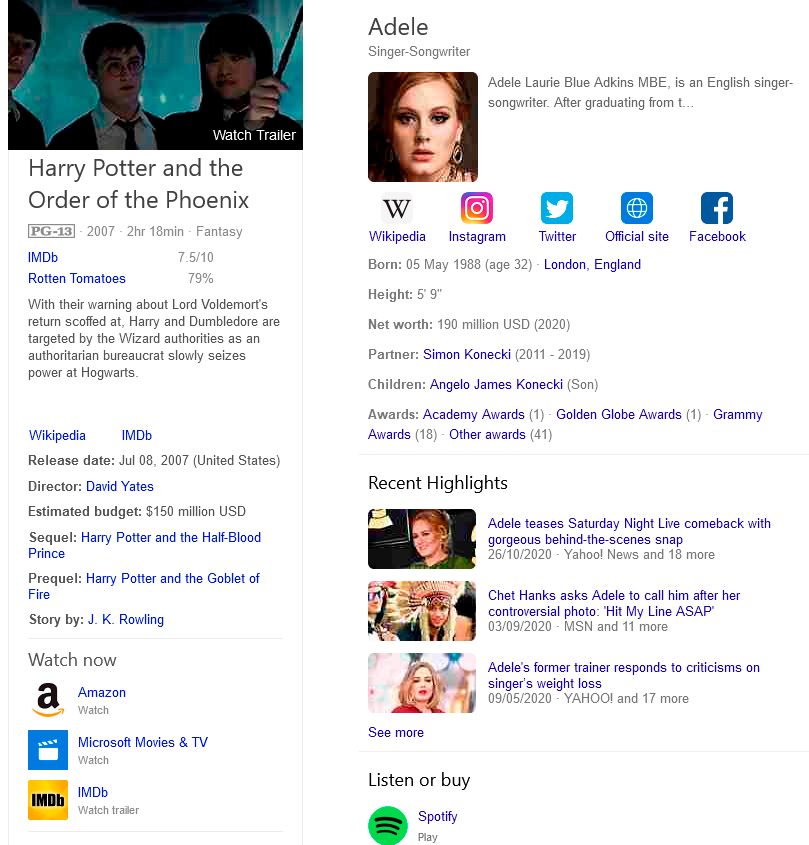}
    \caption{Bing's Knowledge Panel from 2016 (left) and 2020 (right)}
    \label{fig:kpbing}

\end{figure}

The \textbf{Knowledge Panel}, seen in Figures \ref{fig:kpgoogle} and \ref{fig:kpbing}, is perhaps the highlight of SERP features. It is a dynamic feature that provides direct information in various formats within the same panel, pointing to related content. The contents range from text to images, ratings, social profiles, factual information, and similar search topics~\cite{Rosu2020}, helping the user to understand a particular subject quickly and facilitating a more in-depth search~\cite{DannySullivan2020}. This element appeared for the first time in 2014 in Google, lasting until now, and one year later in Bing. The latter is highly similar to Google's one. Its positioning is always at the right of the results container. The basic structure consists of a panel with a top thumbnail of the subject, vertically followed by a title, a website link if applicable, a resume paragraph usually by \emph{Wikipedia}, and a structured list of direct information. Google also includes a block of \emph{People also search for}, while Bing has other blocks, in which timelines seem to be included whenever possible. During the following years, Google introduced other content highly dependable on the search topic and variable in coverage and quality~\cite{Lurie2018}. It is common to have this panel populated by Wikipedia information~\cite{Vincent2021}.

\begin{figure}[h]
    \centering
    \includegraphics[width=1.0\columnwidth]{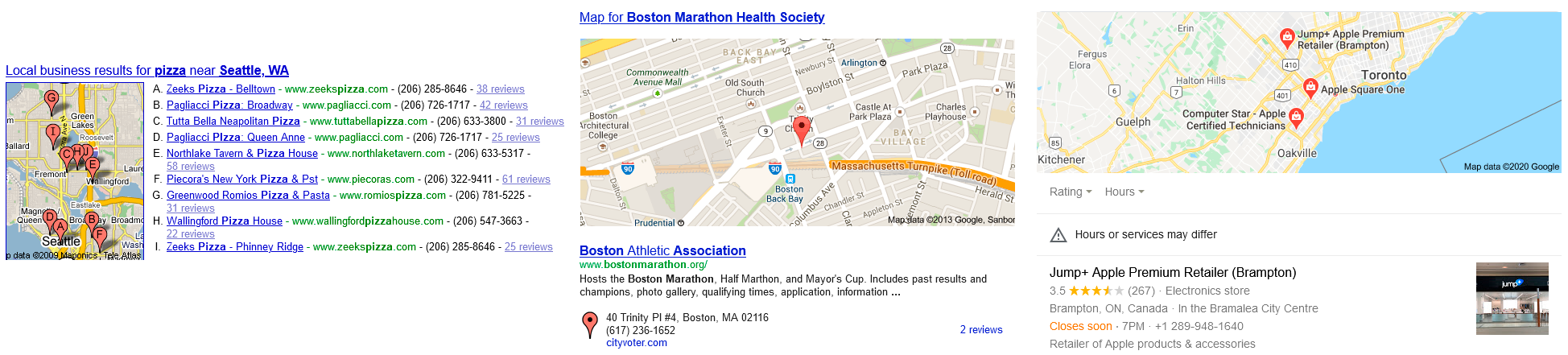}
    \caption{Google's Local Pack from 2009 (left), 2013 (center) and 2020 (right)}
    \label{fig:mapsresults}
\end{figure}

\begin{figure}[h]
    \centering
    \includegraphics[width=0.7\columnwidth]{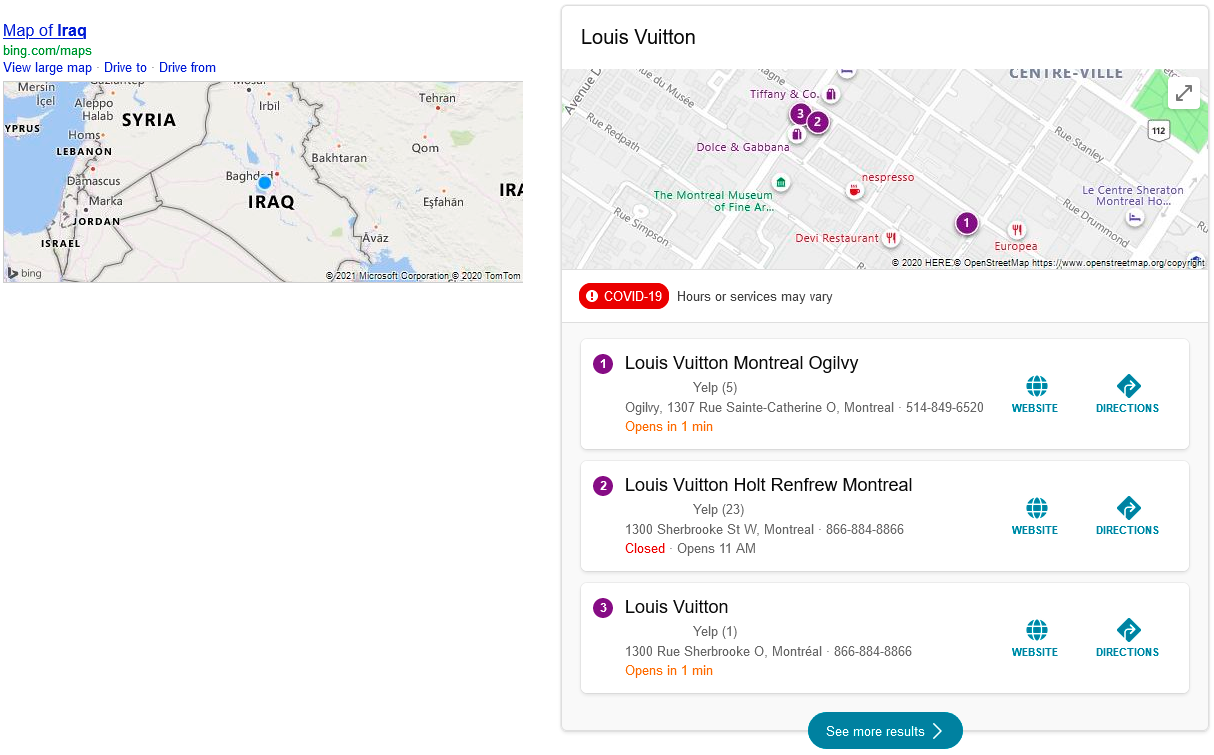}
    \caption{Bing's Local Pack from 2014 (left) and 2020 (right)}
    \label{fig:binglocalp}
\end{figure}

The \textbf{Local Pack}, seen in Figures \ref{fig:mapsresults} and \ref{fig:binglocalp}, is activated by geographical searches, including the main related results from a map application such as \emph{Google Maps}. This element appeared for the first time in 2009, in Google, and scarcely before 2018, when it started to occur more frequently. There are presences of Bing's Local Pack in 2014 and 2020, whose element is similar. Its positioning is highly variable throughout the results container, with the tendency to be in the visible area. The primary content structure consists of a title relating a map to the search query and a list of close locations. Each result includes a map preview with its location and additional information that, in the beginning, was a title, website, phone number, and link to reviews. In 2013, a description, the complete location, and the \emph{similar \& cached} links, like organic results, were added. In 2016, Google introduced options to filter results by rating and schedule. Each result stated if the place was open or closed and had buttons for the place's website or to get its directions, while in 2019, when applicable, a thumbnail image substituted those buttons. When focused on experience locations, like restaurants and hotels, it lists information about prices and testimonials. The map was initially a small image at the left of the results list. The image was later enlarged to full width, and, in 2016, it changed from a static image to an embedded \emph{Google Maps} instance, like later in Bing's case. 

\begin{figure}[h]
    \centering
    \includegraphics[width=0.8\columnwidth]{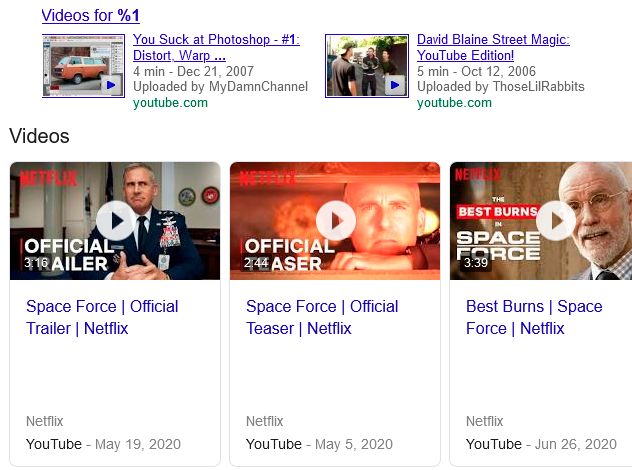}
    \caption{Google's Video Pack from 2010 (top) and 2020 (bottom)}
    \label{fig:videopack}
\end{figure}

\begin{figure}[h]
    \centering
    \includegraphics[width=0.7\columnwidth]{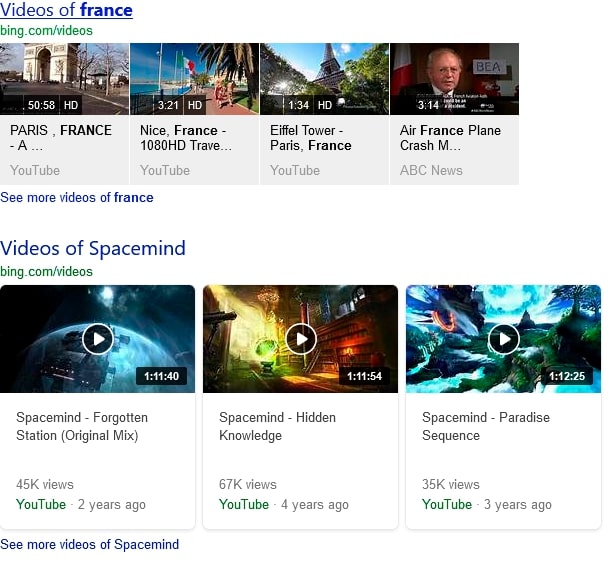}
    \caption{Bing's Video Pack from 2015 (left) and 2019 (right)}
    \label{fig:bingvideos}
\end{figure}

The Video Pack, seen in Figures~\ref{fig:videopack} and \ref{fig:bingvideos}, features content from video platforms like \emph{YouTube}. This element was present in Google for the first time in 2010, staying until now, except from 2012 to 2014. Bing introduced it in 2014, lasting until now. Its positioning is highly variable throughout the results container. The content in Google started with a title associating videos with the search query and a block of two videos, each accompanied by a thumbnail, title, timestamp, description, and URL. Bing's first layouts included more entries. Google's main layout was introduced in 2015, with three or more videos individually in a carousel of cards. The video duration was included in the thumbnail, and the account name was in the card's body. Bing's structure is very similar, substituting a previous video resolution badge with the number of views above the source. In Bing, the initial design was original, with creased square shapes, while the current one is clearly influenced by Google.

\begin{figure}[h]
    \centering
    \includegraphics[width=1.0\columnwidth]{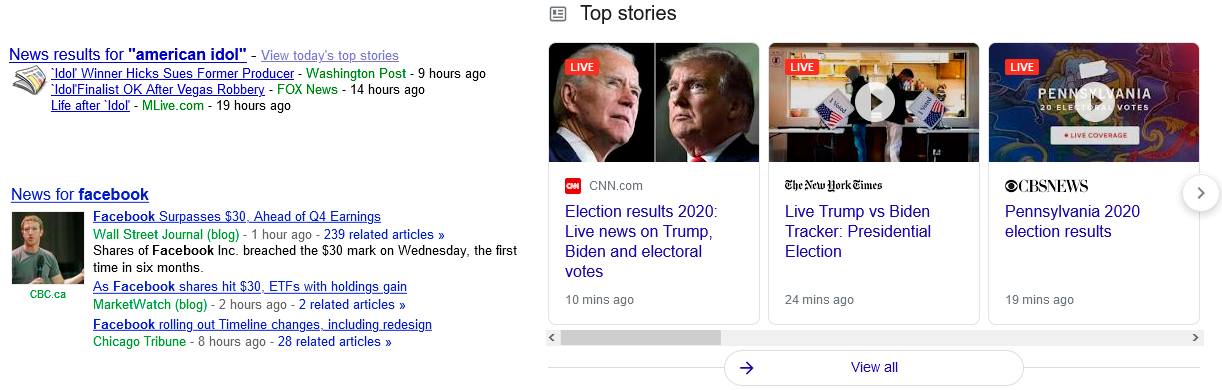}
    \caption{Google's Top Stories from 2006 (left-top), 2013 (left-bottom) and 2020 (right)}
    \label{fig:newsres}
\end{figure}

\begin{figure}[h]
    \centering
    \includegraphics[width=1.0\columnwidth]{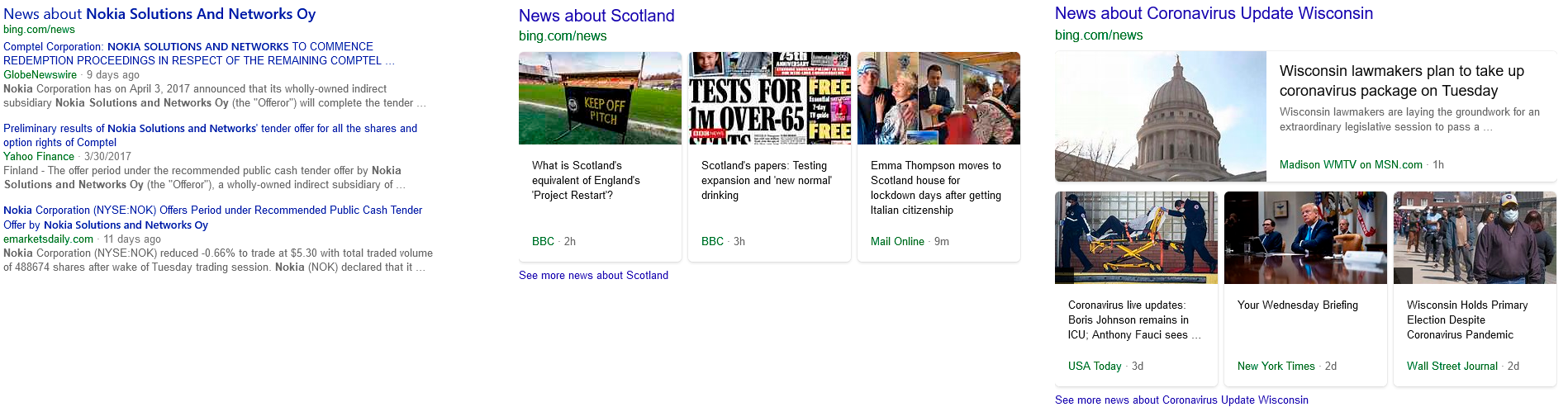}
    \caption{Bing's Top Stories from 2017 (left) and 2020 (center and right)}
    \label{fig:bingnews}
\end{figure}

The \textbf{Top Stories}, seen in Figures \ref{fig:newsres} and \ref{fig:bingnews}, are blocks of three or more recent news considered relevant to the query, recently placed in the form of a carousel~\cite{Rosu2020}. Each story is now presented with a thumbnail, publisher, and timestamp. This element appeared for the first time in Google in 2004, but only after 2011 did it frequently appear, lasting until now. In Bing, we found it in 2010, 2014, and 2017 before a considerable presence in 2020. Its positioning is mainly in the visible area of the results container. The element's content started in Google with a vertical list of news. In 2006, a journal icon was placed at the left of the list, and a link to `today's top stories' was introduced. In 2013, the icon was substituted by a thumbnail for the first news result. In Bing's case, the element is currently more sophisticated. In 2010 it emphasized the first news item. In 2014, this top position started to include a thumbnail. From 2020, two layouts were introduced. One layout is a shorter format, with a horizontal list of three cards. Another arrangement presents a highlight to the main news, in a horizontal card that occupies the entire width, followed by a block with three other news items as in the previous layout and, below them, a block of cards with quotes related to the theme. 

\begin{figure}[h]
    \centering
   \includegraphics[width=1.0\columnwidth]{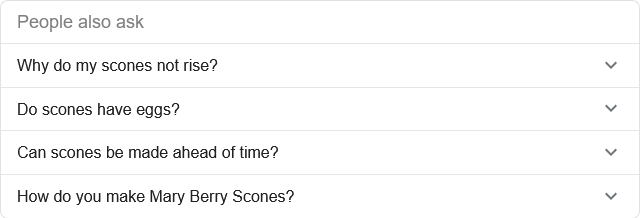}
    \caption{Google's People Also Ask from 2019}
    \label{fig:pplalsoask}
\end{figure}

\begin{figure}[h]
    \centering
    \includegraphics[width=1.0\columnwidth]{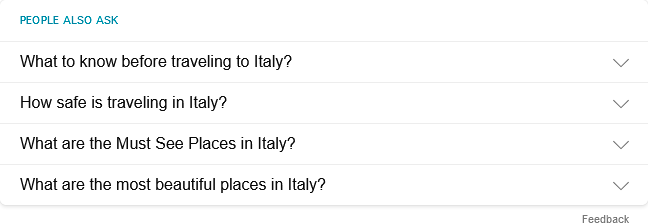}
    \caption{Bing's People Also Ask from 2020}
    \label{fig:bingppl}
\end{figure}

The \textbf{People Also Ask}, seen in Figures \ref{fig:pplalsoask} and \ref{fig:bingppl}, is an accordion of some questions (and answers) suggested by the search engine and related to the search query~\cite{Vaughn2019}. Each expanded element features a featured snippet answering the element's question, complementing the information with the link from where the excerpt is taken~\cite{Moran2019}. This element appeared for the first time in Google in 2016, lasting until now, and was introduced by Bing a half decade later, in 2020. Its positioning is highly variable throughout the results container. No changes can be noticed in content, as the element has kept its shape, four accordion questions, untouchable over time. Bing's element is entirely influenced by Google's one. Studies specifically focused on this feature corroborate more general studies, concluding that searchers issue fewer queries and spend less time on a SERP when the \textit{People also ask feature} is presented~\cite{Pothirattanachaikul2020}. Moreover, users are less likely to interact with a SERP when they first encounter an answer inconsistent with their beliefs~\cite{Pothirattanachaikul2020}.

\begin{figure}[h]
    \centering
    \includegraphics[width=1.0\columnwidth]{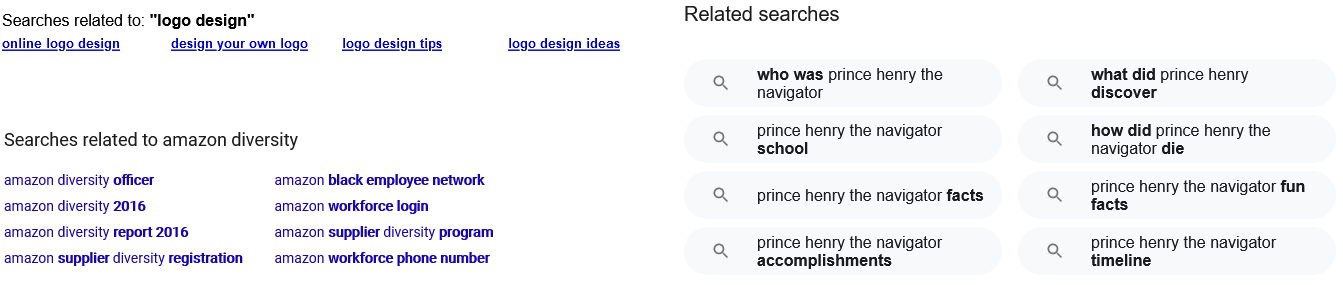}
    \caption{Google's Related Searches from 2008 (left - top), 2017 (left - bottom) and 2020 (right)}
    \label{fig:relsearches}
\end{figure}

\begin{figure}[h]
    \centering
    \includegraphics[width=1.0\columnwidth]{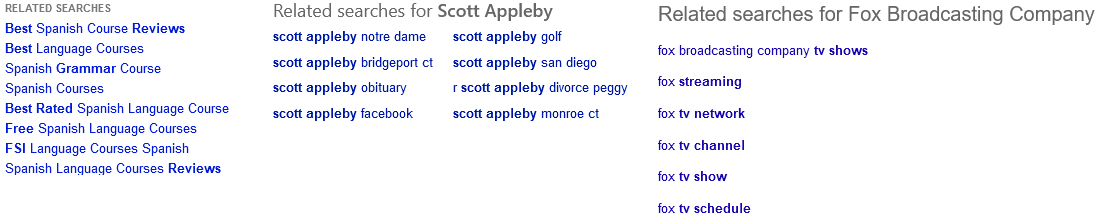}
    \caption{Bing's Related Searches from 2010 (left), 2017 (center) and 2020 (right)}
    \label{fig:bingrelsearches}
\end{figure}

The \textbf{Related Searches}, seen in Figures \ref{fig:relsearches} and \ref{fig:bingrelsearches}, is a common element on SERP pages from a very early age. It offers suggestions for related searches, i.e., queries that are in some way related to the current query and may be good candidates for follow-on queries. These suggestions can be helpful to support exploration or provide query statements that express information needs in different ways~\cite{white2016}. Usually, these suggestions are generated based on search log data, either picking queries that frequently follow the current query~\cite{Jones2006} or clustering queries based on results' clicking~\cite{Craswell2007}. Each link takes the user to the respective SERP. This element appeared for the first time in Google in 2008, lasting until now, except for 2010. In Bing, it was present since the beginning, except for 2011. Its positioning is always at the bottom of the results container. The content differed on how many suggestions would appear and its layout. Each search suggestion is a hyperlinked title pointing to its respective SERP. Initially, it was organized in a matrix of columns. In 2011, Google introduced a two-column layout but results could still be presented in one column if suggestions were verbose. Google applied a search icon to each entry in 2020. Bing showed the two-column layout in 2014, keeping a similar shape until 2019, when items became more spaced for better readability.

\begin{figure}[h]
    \centering
    \includegraphics[width=1.0\columnwidth]{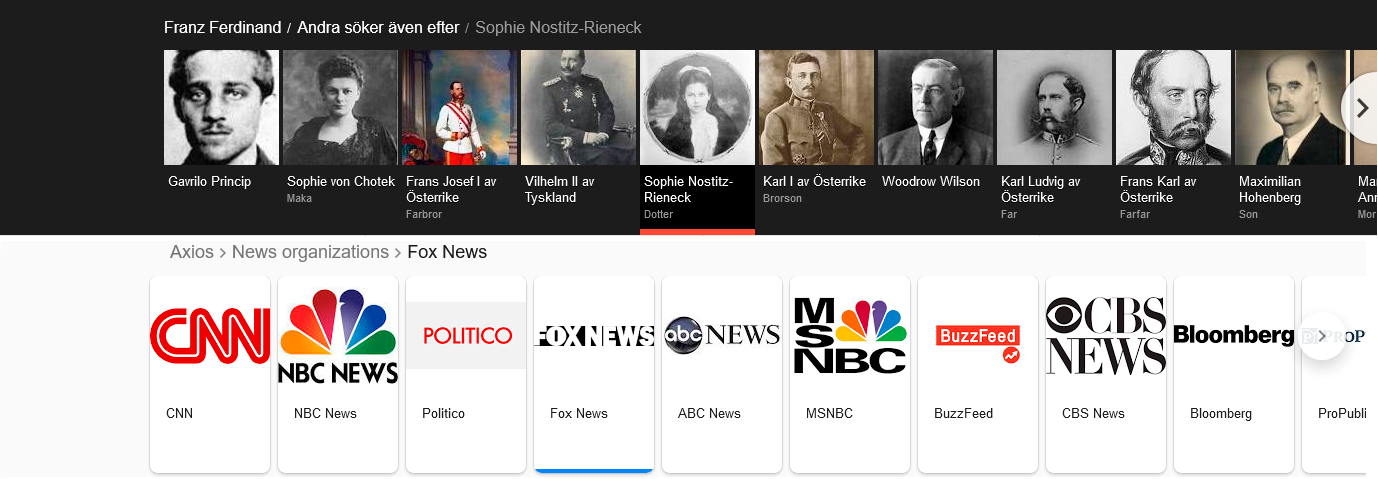}
    \caption{Google's Carousel from 2016 (top) and 2018 (bottom)}
    \label{fig:carousel}
\end{figure}

\begin{figure}[h]
    \centering
    \includegraphics[width=1.0\columnwidth]{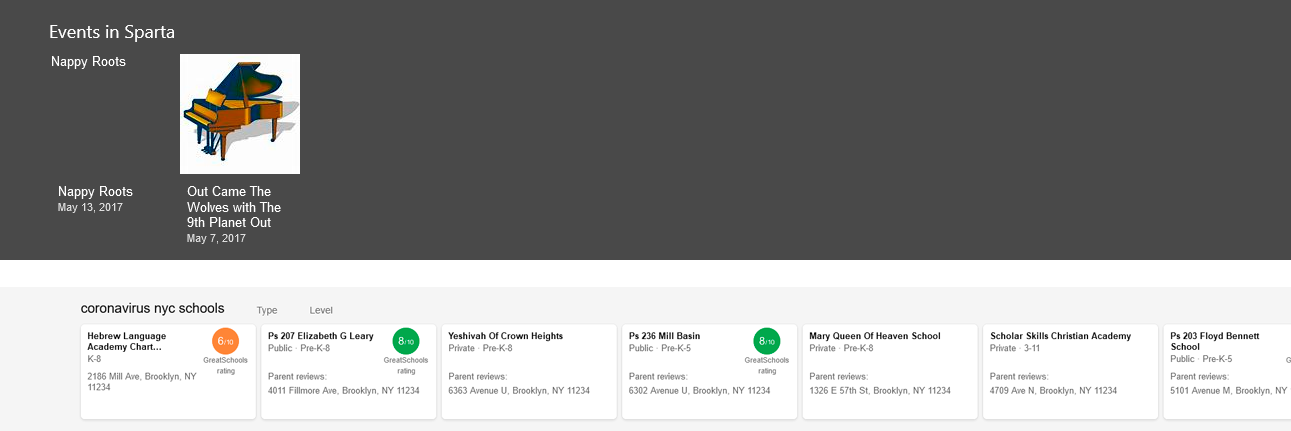}
    \caption{Bing's Carousel from 2017 (top) and 2020 (bottom)}
    \label{fig:bingcarousel}
\end{figure}

A \textbf{Carousel} is a line of several cards, seen in Figures \ref{fig:carousel} and \ref{fig:bingcarousel}, accompanied by a name and highlighting an image, which presents related results at the top of the SERP that are part of the same category. It is possible to scroll through more cards without leaving the SERP. Each card takes the user to the result's respective SERP. This element appeared for the first time in 2015, lasting until 2018 in the dataset, although it is still present today \cite{Rosu2020}. In Bing, it was found twice in the dataset, in 2017 and 2020. Its positioning is always below the search query bar and above the main results container, filling the entire width. Categories were figured as a breadcrumb at the top. Figure \ref{fig:bingcarousel}'s second example of Bing shows a list of schools, representing one of many dynamic possibilities for structuring content, with the school name, type, address, and even a rating if applicable.

\begin{figure}[h]
    \centering
    \includegraphics[width=1.0\columnwidth]{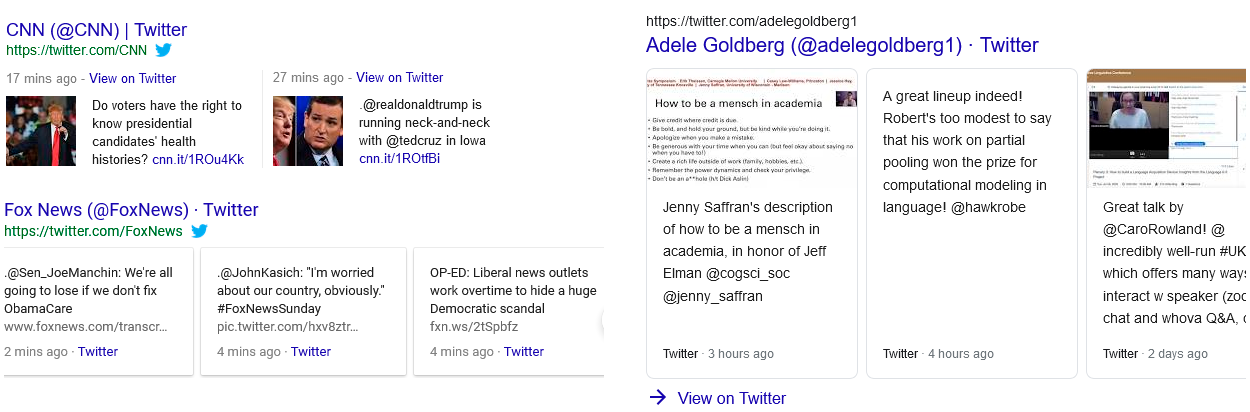}
    \caption{Google's Twitter pack from from 2015 (left - top), 2018 (left - bottom) and 2020 (right)}
    \label{fig:twitterpack}
\end{figure}

\begin{figure}[h]
    \centering
    \includegraphics[width=0.6\columnwidth]{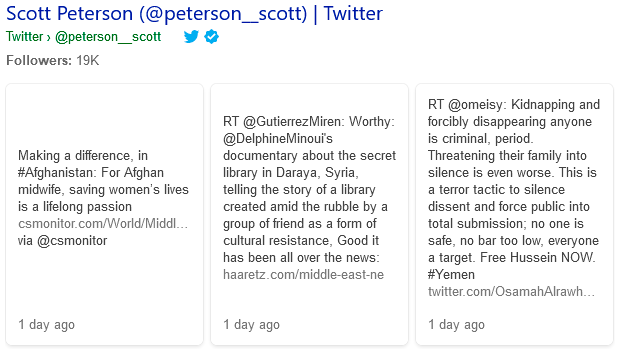}
    \caption{Bing's Twitter Pack from 2019}
    \label{fig:bingtwitter}
\end{figure}

The \textbf{Twitter Pack}, seen in Figures \ref{fig:twitterpack} and \ref{fig:bingtwitter}, presents recent tweets related to the search query. This element appeared for the first time in 2015, lasting until now. Its positioning is always at the bottom of the results container. Its positioning is highly variable throughout the results container. Google's initial structure for content was a header with the Twitter account username and the account URL, followed by two recent results. A thumbnail could accompany a result that always included the tweet, how long it was published, and a link to the tweet's page. In 2017, tweets started to be three or more in a carousel, displayed on individual cards, without showing an image. In 2020, the structure was the same but included a thumbnail for each tweet. Bing's element appears once in the dataset. It displays the person's name, Twitter username, link to the profile, rounded number of followers, and a row of three cards with recent tweets. 

\subsection{Google's exclusive features}
\begin{figure}[h]
    \centering
    \includegraphics[width=1\columnwidth]{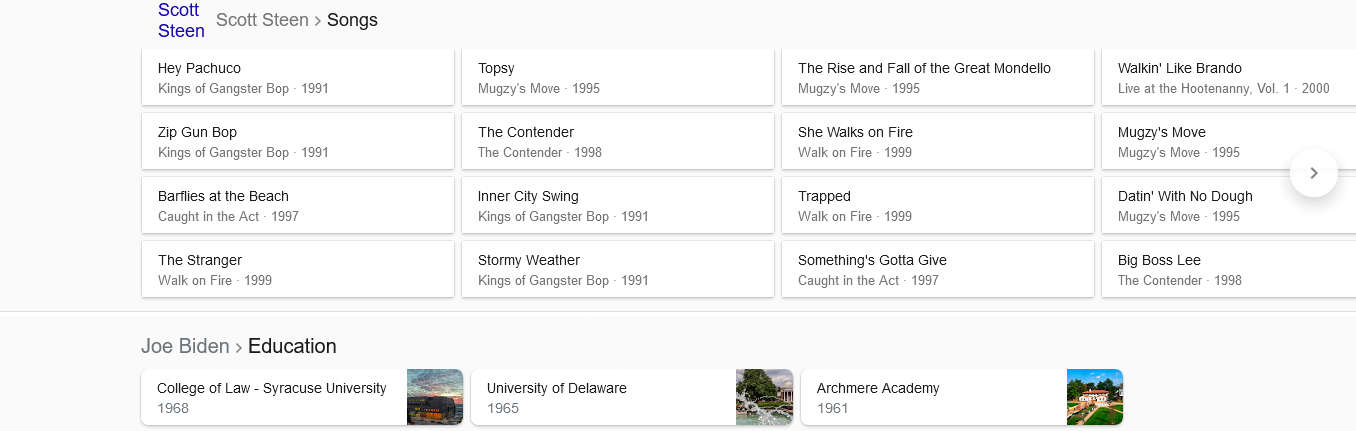}
    \caption{Google's Carousel Grid}
    \label{fig:gartcargrid}
\end{figure}

We found four features exclusive to Google. The \emph{Carousel Grid}, in Figure~\ref{fig:gartcargrid}, is a \emph{Carousel} displaying content in a matrix of cards instead of a line positioned at the same place.

\begin{figure}[h]
    \centering
    \includegraphics[width=1\columnwidth]{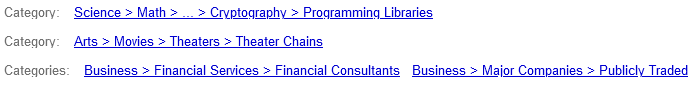}
    \caption{Google's Category Hierarchy}
    \label{fig:gartcategories200.png}
\end{figure}

The \emph{Category Hierarchy}, in Figure~\ref{fig:gartcategories200.png}, is a no longer present element that stated in breadcrumbs one or more categories related to the query. 

\begin{figure}[h]
    \centering
    \includegraphics[width=1\columnwidth]{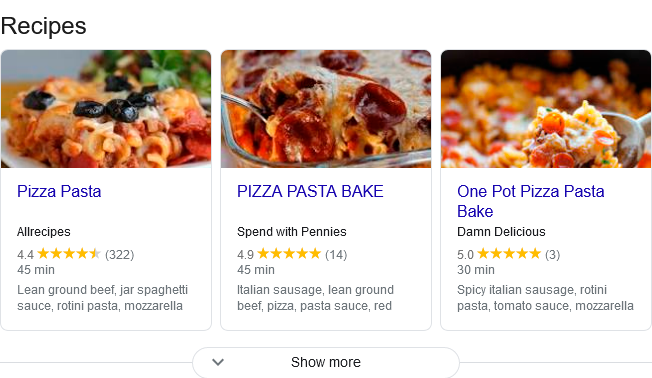}
    \caption{Google's Recipe Cards}
    \label{fig:despatdisclosure.png}
\end{figure}

\emph{Recipe Cards}, in Figure~\ref{fig:despatdisclosure.png}, is a group of cards with links to culinary recipes enriched with ratings, a brief snippet of ingredients, and relevant, concise information.  

\begin{figure}[h]
    \centering
    \includegraphics[width=0.6\columnwidth]{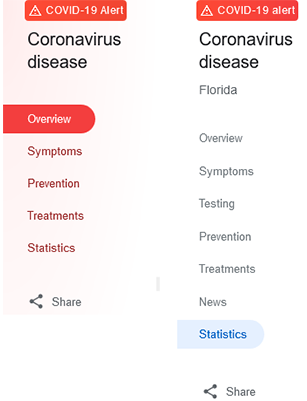}
    \caption{Google's Covid-19 Left Panel}
    \label{fig:gartcovidpanel}
\end{figure}

The \emph{Covid-19 Left Panel}, in Figure~\ref{fig:gartcovidpanel}, allows the user to consult various tabs of updated information regarding Covid-19 panorama and prevention.

\subsection{Bing's exclusive features}
\begin{figure}[h]
    \centering
    \includegraphics[width=1\columnwidth]{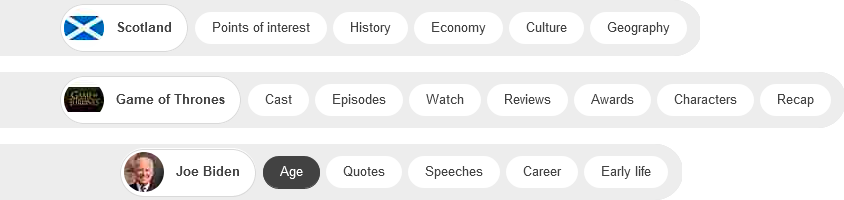}
    \caption{Bing's Categories Bar}
    \label{fig:BCategoriesBar}
\end{figure}

We found three features exclusive to Bing. The \emph{Categories Bar}, in Figure~\ref{fig:BCategoriesBar}, presents a row of buttons referring to query-related categories.

\begin{figure}[H]
    \centering
    \includegraphics[width=1\columnwidth]{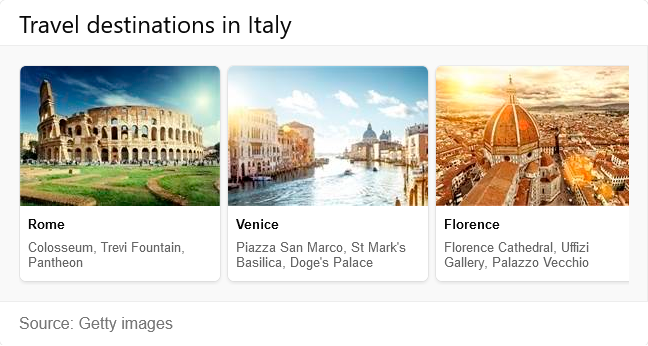}
    \caption{Bing's Travel Destinations}
    \label{fig:BTravelDests}
\end{figure}

The \emph{Travel Destinations} feature in Figure~\ref{fig:BTravelDests} presents a set of three cards with main travel destinations when the query is focused on a country.

\begin{figure}[h]
    \centering
    \includegraphics[width=1\columnwidth]{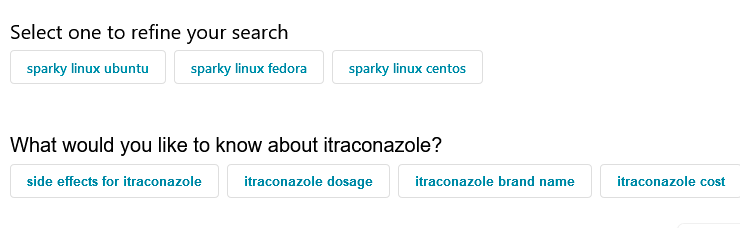}
    \caption{Bing's Search Refinement}
    \label{fig:BRefineSearch}
\end{figure}

The \emph{Search Refinement}, in Figure~\ref{fig:BRefineSearch}, provides query-refinement options to the user.

\section{Discussion and Conclusion} \label{sec:conclusion}
We examined how SERP user interfaces evolved over two decades of presenting search results, using Google and Bing as case studies. We have updated and improved the analysis with an evolution perspective, addressing and analyzing the most sophisticated SERP elements: the features.

We showed that SERP are becoming more diverse in terms of features, aggregating content from different verticals and including more features that provide direct answers, keeping the user on the results page. 

We compared both search engines' SERP and saw that, despite some innovation, expressed on exclusive elements, the layout of both interfaces is similar for the various features analyzed with an almost identical number of SERP features. Besides appearing earlier, Google's features have, throughout its history, also influenced its competitor, Bing, whose similar features would appear after.

\begin{acks}
The Master in Informatics and Computing Engineering and the Department of Informatics Engineering of the Faculty of Engineering of the University of Porto supported this work by funding the registration fee.
\end{acks}

\bibliographystyle{ACM-Reference-Format}
\bibliography{diss-bibliography}

\end{document}